\documentclass[aps,prl,reprint]{revtex4-1}
\usepackage{amsmath}
\usepackage{amssymb}
\usepackage{graphicx}
\usepackage{hyperref}
\usepackage{url}
\usepackage{color,xcolor}
\usepackage{epstopdf}

\begin{document}

\title{Exchange striction driven magnetodielectric effect and potential photovoltaic effect in polar CaOFeS}
\author{Yang Zhang}
\author{Lingfang Lin}
\author{Jun-Jie Zhang}
\author{Xin Huang}
\author{Ming An}
\author{Shuai Dong}
\email{Corresponding author: sdong@seu.edu.cn}
\affiliation{School of Physics, Southeast University, Nanjing 211189, China}
\date{\today}

\begin{abstract}
CaOFeS is a semiconducting oxysulfide with polar layered triangular structure. Here a comprehensive theoretical study has been performed to reveal its physical properties, including magnetism, electronic structure, phase transition, magnetodielectric effect, as well as optical absorption. Our calculations confirm the Ising-like G-type antiferromagnetic ground state driven by the next-nearest neighbor exchanges, which breaks the trigonal symmetry and is responsible for the magnetodielectric effect driven by exchange striction. In addition, a large coefficient of visible light absorption is predicted, which leads to promising photovoltaic effect with the maximum light-to-electricity energy conversion efficiency up to $24.2\%$.
\end{abstract}

\maketitle

\section{Introduction}
The discovery of unconventional superconductivity in fluorine doped LaOFeAs ~\cite{Kamihara:Jacs} with transition temperature $T_{\rm C}=26$ K had stimulate great interests of iron pnictides/chalcogenides ~\cite{Dai:Np,Dagotto:Rmp,Dai:Rmp,Stewart:Rmp,Lumsden:Jpcm}. In the iron pnictide and chalcogenide families, the so-called 1111-series is an important branch, which contains the first-concerned LaOFeAs \cite{Kamihara:Jacs} and the highest-$T_{\rm C}$ bulks \cite{Ren:Cpl,Wang:epl}. Generally, the 1111-series owns layered Fe square lattice, which undergoes a tetragonal-to-orthorhombic structural transitions followed by the stripe antiferromagnetic (AFM) transition ~\cite{Cruz:Nat,Dai:Rmp}.

\begin{figure}
\centering
\includegraphics[width=0.48\textwidth]{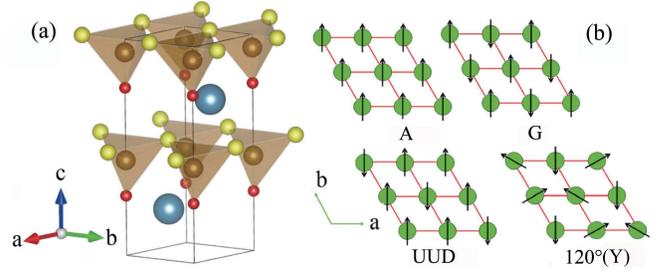}
\caption{(a) Schematic crystal structure of CaFeSO. Blue: Ca; red: O; yellow: S; brown: Fe. (b) Sketch of  possible spin configurations (denoted by arrows) in 2D triangular lattice. Between layers, both the parallel and antiparallel configurations have been calculated.}
\label{Fig1}
\end{figure}

Recently, some other 1111-type transition metal oxysulfides with different structure have been reported. For example, CaO$M$S ($M$=Fe, Zn) forms layered triangular lattice ~\cite{Selivanov:IM,Sambrook:IC,Jin:prb,Delacotte:IC}. As sketched in Fig.~\ref{Fig1}(a), CaOFeS owns a hexagonal structure, whose space group is $P6_3mc$ (No. 186). In each unit cell, there are two $ab$-plane Fe layers, which are built by triangles of O-Fe-S$_3$ tetrahedra. Ca ions intercalate between S and O layers. Although CaOFeS was synthesized more than ten years ago ~\cite{Selivanov:IM}, its physical properties have not been carefully studied until recent years ~\cite{Jin:prb,Delacotte:IC}.

Different from square lattice, triangular lattice provides the geometry for AFM frustration ~\cite{Ramirez:Arms,Collins:CJP,Dong:ap}. For Heisenberg spins, a typical Y-type ground state usually appears with nearest-neighbor spins arranged with $120^{\circ}$ in the two-dimensional (2D) triangular lattice (see Fig.~\ref{Fig1}(b)), for example in Sr$_3$NiTa$_2$O$_9$ \cite{Liu:IC}, Ba$_3$MnNb$_2$O$_9$ \cite{Lee:Prb14.2}, $A$CrO$_2$ \cite{Seki:Prl08}, RbFe(MoO$_4$)$_2$ \cite{Kenzelmann:Prl07,Hearmon:Prl}, and hexagonal $R$MnO$_3$ \cite{Lin:Prb16}. Interestingly, such Y-type magnetism, with noncollinear spin pairs, can lead to multiferroicity in some compounds ~\cite{Dong:ap,Liu:IC,Lee:Prb14.2,Kenzelmann:Prl07,Hearmon:Prl,Seki:Prl08}. While in the Ising-spin limit, spins arranged in 2D triangular lattice can also form some exotic patterns \cite{Yao:prb06.01,Yao:prb06.02,Yao:prb07}. For CaOFeS, the neutron experiment at $6$ K identified an Ising stripe AFM order, i.e. the so-called G-type AFM ~\cite{Jin:prb}. In addition, Delacotte \textit{et al.} reported a prominent magnetodielectric effect near the N\'eel temperature ~\cite{Delacotte:IC}.

Considering these experimental advances, there are several interesting physical questions. First, it is surprising that the magnetic order is the collinear G-type AFM, instead of the Y-type noncollinear one. A possible reason is that spin itself is the Ising type instead of the Heisenberg type. However, the $3d^6$ electron configuration of Fe does not own strong spin-orbit coupling (SOC), which can not lead to strong single-axis magnetocrystalline anisotropy. Second, what's the origin of the magnetodielectric effect? Recently, some multiferroic iron selenides, e.g. BaFe$_2$Se$_3$ and KFe$_2$Se$_2$, were predicted ~\cite{Dong:PRL14,Zhang:rrl}. Is CaOFeS one more multiferroic member in the iron selenide family? In fact, the structure of CaOFeS is polar due to the unequivalence of S and O. However, only polar structure is not sufficient for ferroelectricity, considering the non-reversible positions of S and O.

Furthermore, it is recently suggested that ferroelectric materials can promote the efficiency of photovoltaic effects, since the internal electric field from the spontaneous electric polarization can help the separation of photogenerated electrons/holes ~\cite{Butler:EES,Yuan:Nm}. For example, the power conversion efficiency of Bi$_2$FeCrO$_6$ was found to reach $8.1\%$ ~\cite{Nechache:Npho} and the theoretical upper-limit of efficiency for ferroelectric hexagonal TbMnO$_3$ was predicted to be $\sim33\%$ \cite{Huang:Prb}. In fact, the polar structure, as in CaOFeS, even without ferroelectricity, breaks the spatial inversion symmetry, can lead to similar function to separate photon-generated electrons/holes.

In the present work, the magnetic properties, electronic structure, magnetodielectric effect, and optical properties will be theoretically investigated. On one hand, the collinear G-type AFM ground state, as well as magnetodielectric effect, have been verified and well explained. On the other hand, the excellent visible optical absorption has been predicted for CaOFeS, which leads to potential prominent photovoltaic effect.

\section{Methods}
The first-principles electronic structure calculations are performed using the Vienna {\it ab initio} Simulation Package (VASP) with the projector augmented-wave (PAW) potentials. In the present study, different exchange functions (LDA, GGA-PBE, and GGA-PBEsol) have been tested. The GGA-PBE function can give best description of crystal structure of CaOFeS, and thus will be adopted in the following calculations ~\cite{Kresse:Prb,Kresse:Prb96,Blochl:Prb,Perdew:Prl}. The Hubbard $U_{\rm eff}$($=U-J$) is imposed on Fe's $d$ orbitals using the Dudarev implementation ~\cite{Dudarev:Prb}. Different values of $U_{\rm eff}$ are tested in the range $0$ eV to $4$ eV, considering the weak to intermediate strength of the Fe correlation effects in these systems ~\cite{Qazilbash:np,Dai:PNAS}.

To accommodate the magnetic orders, various possible magnetic structures are considered for Fe lattice, as shown in Fig.~\ref{Fig1}(b). In our calculation, the plane-wave cutoff is $550$ eV. The $k$-point mesh is $9\times5\times3$ for the G-type AFM and A-type AFM, which is accordingly adapted for the UUD-type AFM and $120^{\circ}$ Y-type AFM. Both the lattice constants and atomic positions are fully relaxed until the force on each atom is below $0.01$ eV/{\AA}.

To calculate the optical properties precisely, the hybrid functional calculation based on the Heyd-Scuseria-Ernzerhof (HSE06) exchange has been adopted \cite{Heyd:Jcp,Heyd:Jcp04,Heyd:Jcp06}.

Besides DFT, the Markov chain Monte Carlo (MC) method with the Metropolis algorithm is employed to simulate the magnetic phase transition in a $18\times18\times6$ lattice with periodic boundary conditions. In our MC simulation, the first $1\times10^4$ MC steps (MCSs) are used for thermal equilibrium, then the following $1\times10^4$ MCSs are used for measurements. In all simulated temperatures ($T$), the acceptance ratio of MC updates is controlled to be about $50\%$ by adjusting the updating windows for spin vectors. The quenching process (i.e. gradually cooling from high $T$ to low $T$) is adopted in our MC simulation. Specific heats per site ($C_v$) and the spin structure factor ($S$($\textbf{k}$)) are measured as a function of $T$ \cite{Dong:Prb08,Dong:Prb08.2,Lin:CPB,Lin:FOP}.

The photovoltaic energy conversion efficiency is calculated using a spectroscopic limited maximum efficiency method ~\cite{Yu:prl12}. The photovoltaic energy conversion efficiency can be calculated as:
\begin{equation}
\eta=P_{\rm out}^{\rm max}/P_{\rm in}
\end{equation}

Here, $P_{\rm out}^{\rm max}$ is the maximum electrical output power density and $P_{\rm in}$ is the total incident solar power density. The numerically maximum $P_{\rm out}$ could be calculated from \cite{Green:book}:
\begin{equation}
P_{\rm out}=JV=(J_{\rm SC}-J_{\rm D})V=[J_{\rm SC}-J_0(e^{\frac{qV}{kT}}-1)]V
\label{Eq2}
\end{equation}
Here, a photovoltaic cell was approximated as an equivalent ideal diode illuminated. $J_{\rm SC}$ is the short-circuit current density under illumination, which can be obtained from $J_{\rm SC}=e\int_{0}^{\infty }a(E)I_{\rm sun}(E)dE$, where $a(E)$ is the phonton absorpitivity and $I_{\rm sun}(E)$ is the solar radiation flux. Furthermore, the $a(E)$ can be described as $a(E)=1-e^{-\alpha(E)L}$, where depends on the absorption coefficient $\alpha(E)$ and the thickness $L$. $J_{\rm D}$ is the dark current density which depends on the electron-hole recombination current density. $J_0$ is the reverse saturation current density which involving nonradiative part $J_{0}^{\rm nr}$ and radiative part $J_{0}^{\rm r}$ at temperature $T$ and voltage $V$. The reverse saturation current density $J_0$ can be calculated as:
\begin{equation}
J_0=J_{0}^{\rm nr}+J_{0}^{\rm r}=J_{0}^{\rm r}/f_{\rm r}
\end{equation}
$f_{\rm r}$ is the fraction of the radiative electron-hole recombination current, which can be roughly estimated as \cite{Yu:prl12}:
\begin{equation}
f_{\rm r}=e^{-\varDelta/kT}
\end{equation}
where $k$ is the Boltzmann constant and $\varDelta=E_{\rm g}^{\rm da}-E_{\rm g}$ near absorption threshold of a pure semiconductor. Here, $E_{\rm g}^{\rm da}$ is electric-dipole allowed direct band gap and $E_{\rm g}$ is indirect-band gap \cite{Yu:prl12}.

It is well known that the nonradiative recombination is important for materials with indirect band gap. Here, the nonradiative recombination is considered to calculate $f_r$. $\varDelta$ can be obtained from the band structure. In this approximation, the radiative recombination rate can be obtained as a detailed-balance principle that the rates of emission and absorption through cell surfaces should be equal \cite{Tiedje:ITED}. Hence, the current $J_{0}^{\rm r}$ can be calculated by the black-body radiation absorption: $J_{0}^{\rm r}=e\int_{0}^{\infty}a(E)I_{\rm bb}(E)dE$, where $I_{\rm bb}$ is the black-body radiation flux. Finally, the $J_{\rm SC}$ and $J_0$ can be obtained once the absorption coefficient is calculated, and then the photovolatic energy conversion efficiency can be predicted with Eq.~\ref{Eq2}.

\section{Results \& discussion}
\subsection{Magnetism and electronic structure}
\begin{figure}
\centering
\includegraphics[width=0.48\textwidth]{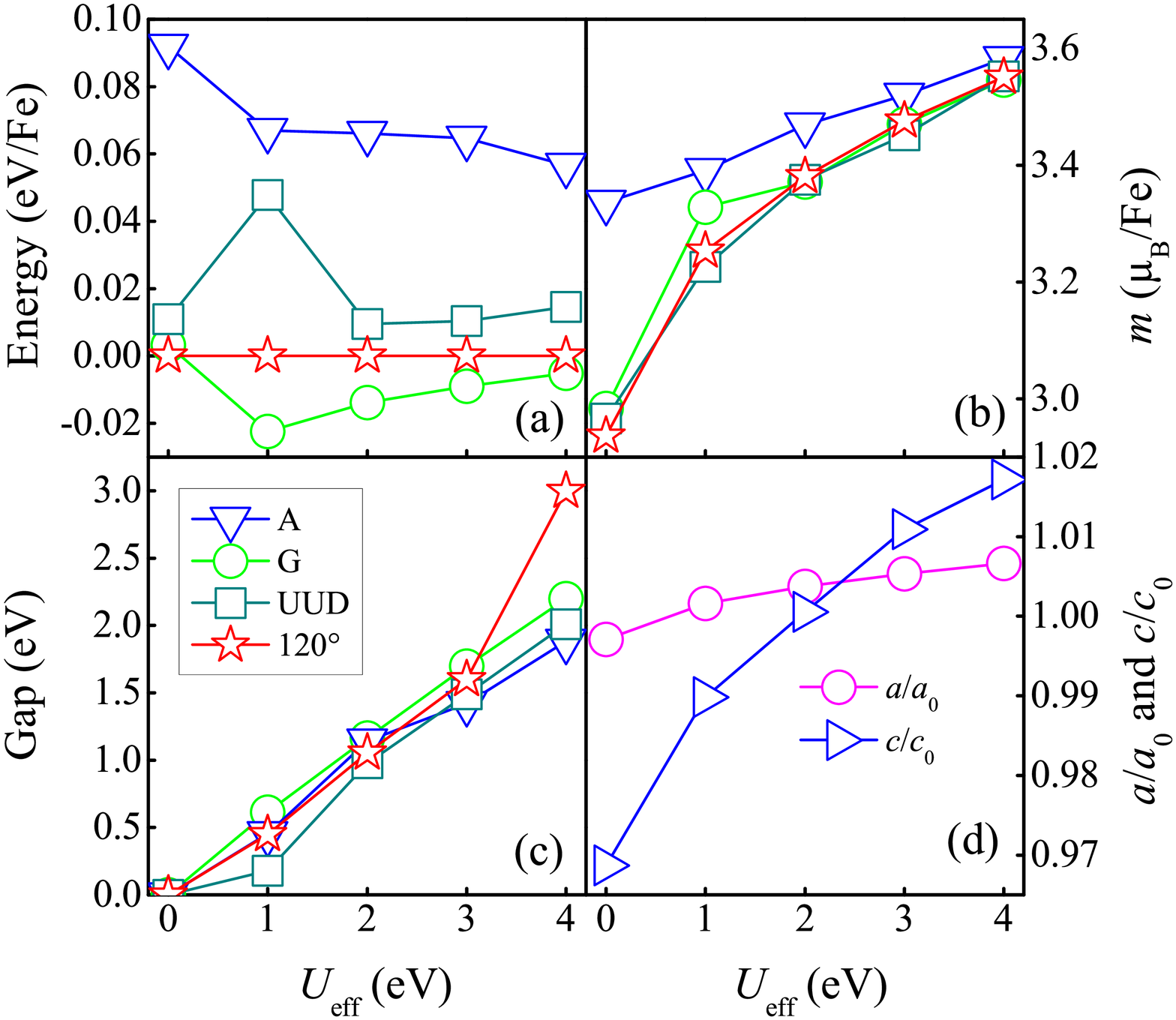}
\caption{DFT results of CaFeSO as a function of $U_{\rm eff}$. (a) Energy (per Fe) of various magnetic orders. The A-AFM state is taken as the reference. (b) Local magnetic moment of Fe calculated within the default Wigner-Seitz sphere. (c) Band gaps. (d) Relaxed Lattice constants, normalized to the experimental ones. }
\label{Fig2}
\end{figure}

First, the energies of various magnetic orders for relaxed structures are summarized in Fig.~\ref{Fig2}(a) as a function of $U_{\rm eff}$. The G-AFM state always own the lowest energy among all candidate configurations except in the pure GGA limit ($U_{\rm eff}=0$), in agreement with the neutron experiments. The spin-orbit coupling (SOC) effect has also been considered (not shown here), which does not alter the conclusion.

Second, the calculated local magnetic moment per Fe is displayed in Fig.~\ref{Fig2}(b). With increasing $U_{\rm eff}$, the moment of each Fe in G-AFM state increases from $2.98$ $\mu_{\rm B}$/Fe to $3.58$ $\mu_{\rm B}$/Fe, which is slightly higher than those obtained in neutron experiments. The overestimated local moment in DFT calculation is widely existing for iron-based pnictides/chalcogenides, which may be related to the itinerant property of electrons in these materials ~\cite{Yin:Nm10,Hansmann:prl10,Mazin:prb,Mazin:np,Zhang:rrl,Zhang:Prb17}.

Third, the band gaps are displayed in Fig.~\ref{Fig2}(c). All magnetic ordered states are insulating (except at $U_{\rm eff}=0$) and the gaps increase with $U_{\rm eff}$, as expected. Noting the experimental gap fitting from resistivity is $0.21$ eV ~\cite{Jin:prb}, which is usually underestimated comparing with the intrinsic one.

Fourth, the normalization of optimized lattice constants are displayed in Fig.~\ref{Fig2}(d). It is clear that $U_{\rm eff}=2$ eV can give the best accurate structure.

According to these results, we can conclude that a nonzero $U_{\rm eff}$ is necessary. In the following, the $U_{\rm eff}=2$ eV will be adopted by default, if not noted explicitly. In fact, our GGA+$U$ ($U_{\rm eff}=2$ eV) calculation leads to the relaxed crystal constants ($a=3.772$ {\AA}, $c=11.378$ {\AA}) for the ground G-AFM state, which is very closed to the neutron experimental results ($a_0=3.752$ {\AA}, $c_0=11.384$ {\AA}) at $6$ K ~\cite{Jin:prb}. Such agreement provides a reliable base for following magnetodielectric study, which may seriously rely on the accurate structure.

\begin{figure}
\centering
\includegraphics[width=0.48\textwidth]{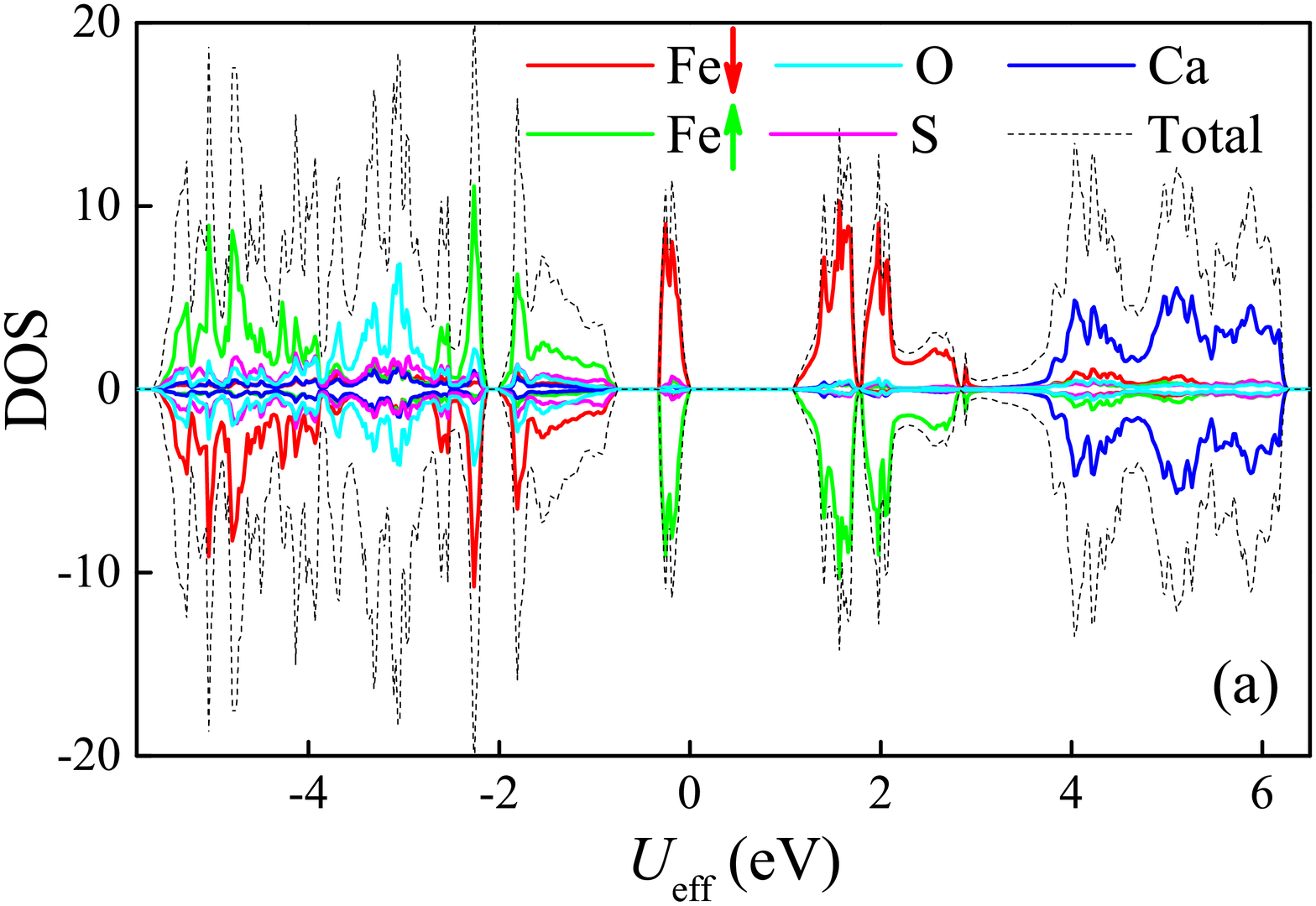}
\includegraphics[width=0.48\textwidth]{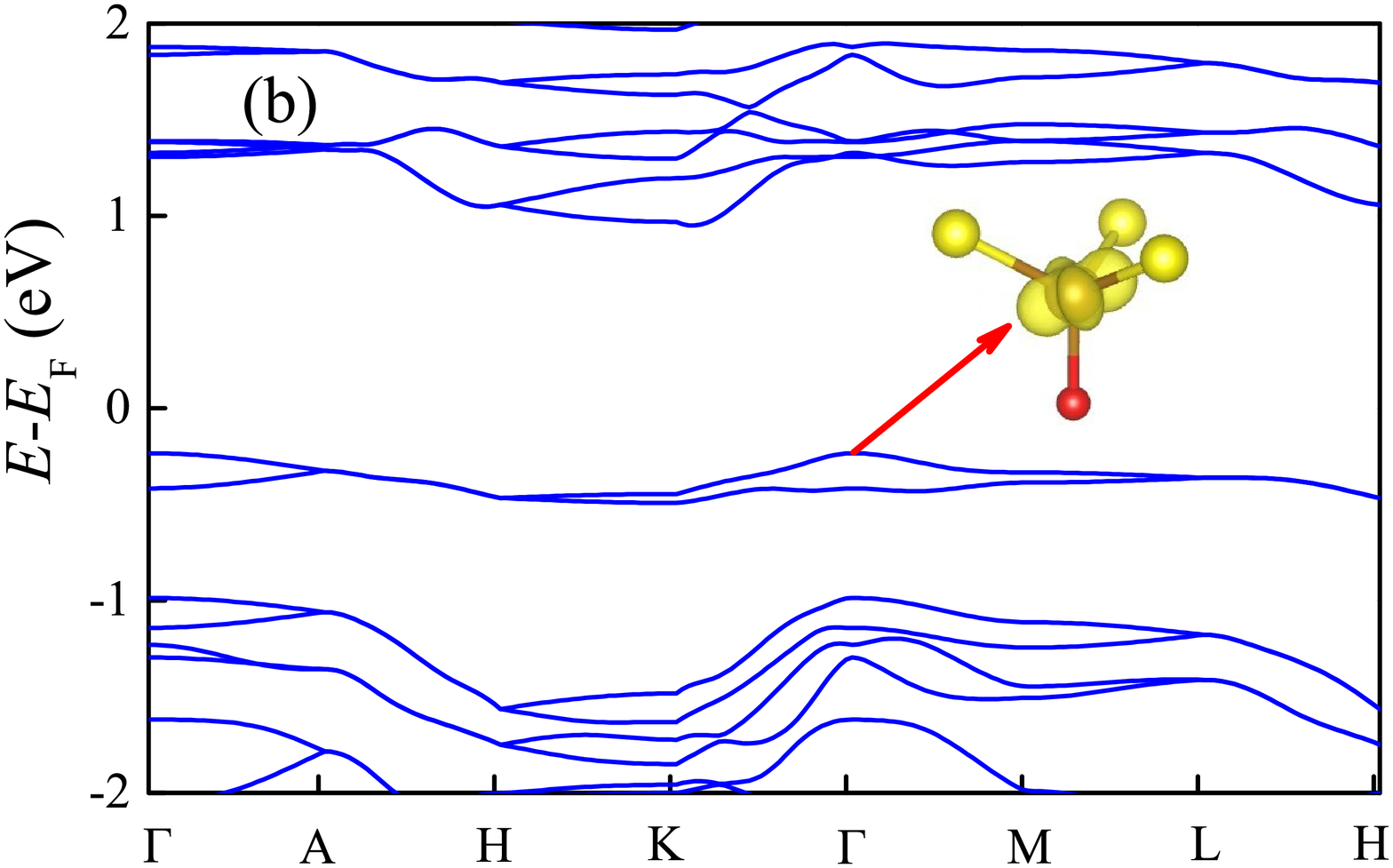}
\includegraphics[width=0.48\textwidth]{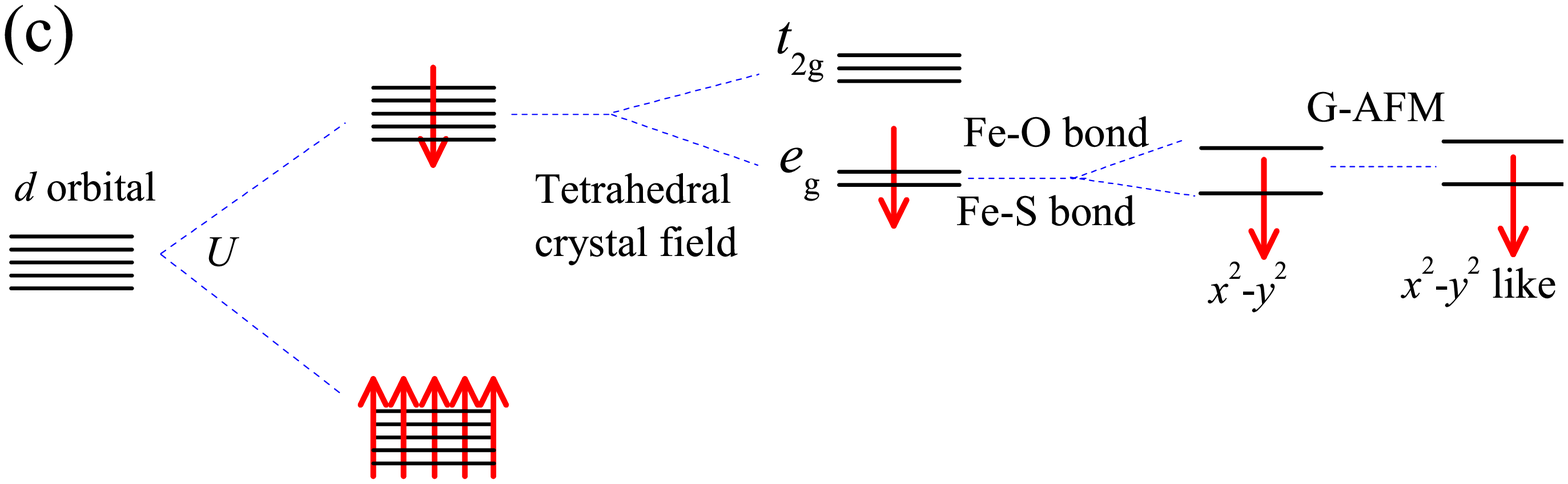}
\caption{Electron structure for G-AFM calculated at $U_{\rm eff}=2$ eV. (a) DOS. Both the total DOS and atomic projected DOS are presented. The Fe($\uparrow$) and Fe($\downarrow$) denote the spin-up/spin-down irons. (b) Band structure near the Fermi level. Insert: the electron density plot of the topmost valence band, which show clear $x^2-y^2$-like characters. (c) The energy splitting of Fe's $3d$ orbitals, which leads to the $x^2-y^2$-like orbital.}
\label{Fig3}
\end{figure}

According to the calculated density of states (DOS) (Fig.~\ref{Fig3}(a)), the bands near the Fermi level are mostly contributed by Fe's $3d$ orbitals and the Fe ion is in the high spin state. Furthermore, the band plot (Fig.~\ref{Fig3}(b)) indicates that CaFeSO is an semiconductor with an indirect band gap about $1.16$ eV. The energy splitting of Fe's $3d$ orbitals is sketched in Fig.~\ref{Fig3}(c). First, the Hubbard replusion generates the Mott splitting between spin-up and spin-down channels. The six $3d$ electrons (per Fe) occupy full spin-up orbitals and one spin-down orbital. Second, the tetrahedral crystal field leads to two low-lying $e_{\rm g}$ orbitals ($x^2-y^2$ and $3z^2-r^2$) and three higher energy $t_{\rm 2g}$ orbitals. Third, the Fe-O bond ($1.867$ {\AA}) is shorter than Fe-S bonds ($>2.3$ {\AA}), which shifts up the energy of orbital $3z^2-r^2$. Thus, the one spin-down electron occupies the $x^2-y^2$ orbital. Furthermore, since the G-type AFM breaks the trigonal symmetry (to be further discussed in the following sub-sections), the lengths of three Fe-S bonds become different: one is $2.337$ {\AA} and other two are $2.406$ {\AA}. As a result, the occupied orbital is slightly distorted from the ideal $x^2-y^2$ one.

\subsection{The antiferromagnetic transition}
Considering the insulating antiferromagnetism, the magnetism of CaFeSO can be described using a Heisenberg model \cite{Jin:prb}:
\begin{eqnarray}
\nonumber H&=&-J_1\sum_{<ij>}\textbf{S}_i\cdot\textbf{S}_j-J_2\sum_{[kl]}\textbf{S}_k\cdot\textbf{S}_l\\
&&-J_3\sum_{\{mn\}}\textbf{S}_m\cdot\textbf{S}_n+A\sum_i(\textbf{S}_i^z)^2.
\end{eqnarray}
where $J_1$/$J_2$ are the in-plane exchange interactions between nearest-neighbor/next-nearest-neighbor spin pairs; $J_3$ is the out-of-plane exchange interaction between the nearest-neighbor spins; $A$ is the single ion magnetic anisotropic coefficient. By comparing the DFT energies of various magnetic states (with the experimental lattice), the coefficients of such a Heisenberg model can be extracted: $J_1=-9.35$ meV, $J_2=-3.03$ meV, and $J_3=-0.50$ meV, respectively. Since the AFM $J_1$ item prefers the Y-type noncollinear order than the G-type collinear one, it is the considerable large $J_2$ determines the G-type AFM over the Y-type AFM. The weak AFM coupling between layers also agrees with the experimental observation \cite{Jin:prb}.

By incorporating the SOC effect, the anisotropic constant $A$ is estimated to be $-1.29$ meV, implying the easy axis is along the $c$ direction, which further enhances the collinear G-type AFM.

\begin{figure}
\centering
\includegraphics[width=0.48\textwidth]{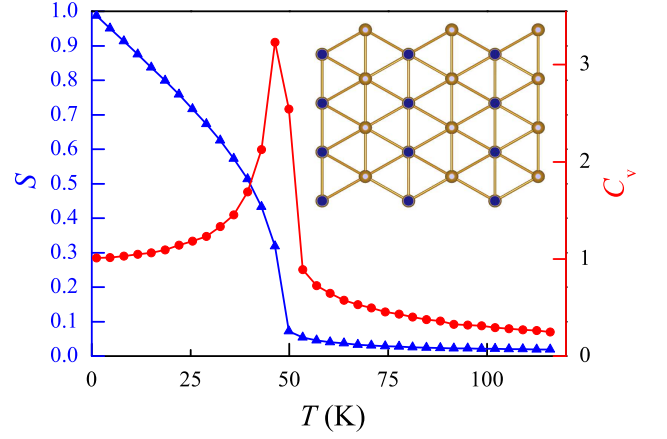}
\caption{MC results for the (in-plane) spin structure factor ($S$) for the G-AFM order and specific heat ($C_v$) as a function of $T$. Inset: the real space spin pattern obtained from low-$T$ MC plus optimization (to further reduce the thermal fluctuation). Different colors denote spin-up and -down along the $c$-axis, while the in-plane components are almost zero.}
\label{Fig4}
\end{figure}

The AFM $J_1$ and considerable $J_2/J_1$ imply the exchange frustration. Although above DFT calculation excluded the $120^\circ$ Y-type AFM as the ground state for CaOFeS, it remains necessary to double-check the ground state since in DFT only a few candidates have been considered. Here, unbiased MC simulation without any preset state is employed to verify the G-AFM, based on the aforementioned coefficients extracted from DFT.

As shown in Fig.~\ref{Fig4}(a), both the spin structure factor and specific heat demonstrate a phase transition at $\sim46$ K, which is close to the experimental value ($\sim36$ K). The spin structure factor, as well as the real space spin order, confirms the G-AFM as the ground state.

\subsection{Exchange striction \& magnetodielectric effect}
Experimentally, the magnetodielectric effect is observed around the AFM $T_{\rm N}$ \cite{Delacotte:IC}, which can also be understood by the following analysis and DFT calculation.

First, the crystalline structure of CaOFeS, with a space group $P6_3mc$ and point group $6mm$, is polar, due to the unequivalence of O and S. But this polar structure is irreversible since the layers of O and S are fixed. Second, the special G-AFM order breaks the trigonal (i.e. $120^{\circ}$ rotation) symmetry of triangular lattice. In each Fe triangle, there are one Fe($\uparrow$)-Fe($\uparrow$) (or Fe($\downarrow$)-Fe($\downarrow$)) bond and two Fe($\uparrow$)-Fe($\downarrow$) bonds, which are no longer symmetric. This breaking of symmetry will distort the lattice, by shrinking the Fe($\uparrow$)-Fe($\downarrow$) bonds but elongating others. According to our DFT optimized structure, such exchange striction will also result in the change of Fe-Fe distance up to $0.008$ {\AA}, i.e. the triangles are no longer regular but with $0.13^{\circ}$ correction for $\angle_{\rm Fe-Fe-Fe}$. Such a tiny distortion is beyond the current experimental precision of structural measurement. Third, the distortion of Fe-S bonds are more serious, reaching $0.069$ {\AA} as mentioned before. It is the displacements of S ions along the $c$-axis responsible to the observed magnetodielectric effect. Qualitatively, these distortions exist independent on the choice of $U_{\rm eff}$. Comparing with the polar distortions in other typical multiferroic materials, e.g. $o$-TbMnO$_3$ or $h$-YMnO$_3$, this amplitude of distortion is not small \cite{Walker:Sci,Aken:Nm}.

\begin{figure}
\centering
\includegraphics[width=0.48\textwidth]{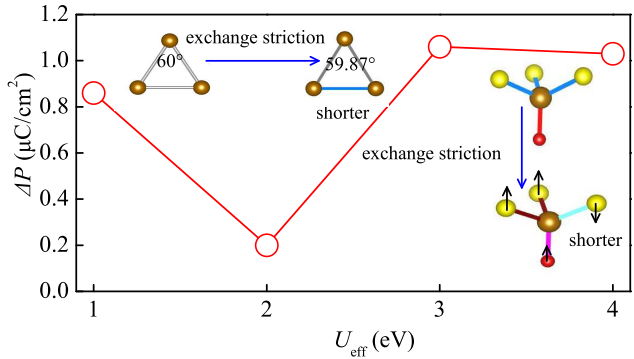}
\caption{The magnetoelectricity, i.e. change of polarization upon G-AFM ordering.}
\label{Fig5}
\end{figure}

The standard Berry phase method ~\cite{King-Smith:Prb,Resta:Rmp} is adopted to estimate the change of dipole moment upon G-AFM ordering. The experimental atomic position is taken as the reference. The change of polarization ($\Delta P$) calculated is along the $c$-axis, which is $1.94$ $\mu$C/cm$^2$ at $U_{\rm eff}=2$ eV and decreasing with $U_{\rm eff}$ (Fig.\ref{Fig5}). Such a $\Delta P$ is considerable large for magnetodielectric effect, reflecting by a dielectric peak around $T_{\rm N}$ as observed in experiment \cite{Delacotte:IC}. In addition, our calculation proposes to perform the pyroelectricity measurement, which should observe a pyroelectric current peak around $T_{\rm N}$.

\subsection{Optical absorption \& photovoltaic effect}
\begin{figure}
\centering
\includegraphics[width=0.48\textwidth]{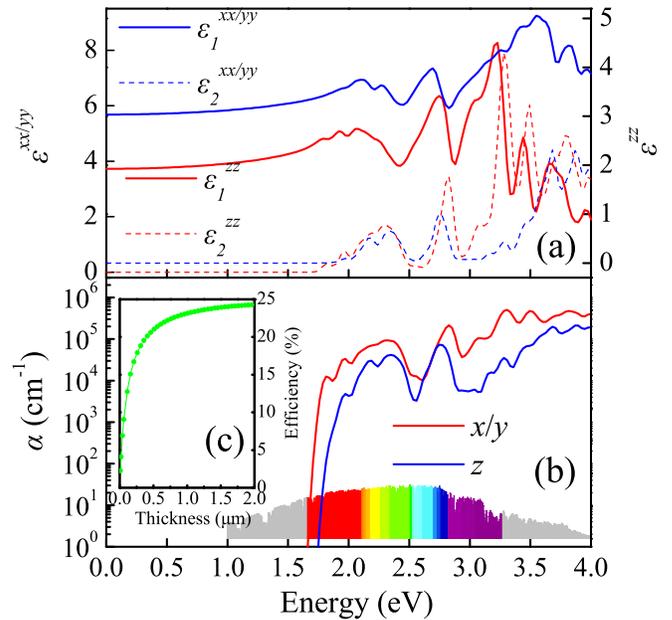}
\caption{Optical properties calculated using HSE06 exchange method. (a) The calculated dielectric spectra. Red: real part; Blue: imaginary part. Solid: $\varepsilon^{xx}$ or $\varepsilon^{yy}$; Broken: $\varepsilon^{zz}$. (b) The calculated absorption coefficient $\alpha(\omega)$ of CaFeSO. The energy spectrum of solar light is shown for reference. (c) Calculated maximum photovoltaic energy conversion efficiency for CaOFeS as a function of absorber layer thickness.}
\label{Fig6}
\end{figure}

As shown in above subsection, the band gap of CaOFeS is $1.16$ eV at $U_{\rm eff}=2$ eV, lower than the visible light lower-limit. However, it is well known that the DFT technique usually underestimates band gaps, although here the correction $U$ has been added. To calculate the optical properties precisely, the hybrid functional calculation based on the Heyd-Scuseria-Ernzerhof (HSE06) exchange has been adopted \cite{Heyd:Jcp,Heyd:Jcp04,Heyd:Jcp06}. According to literature, the fraction of exact exchange coefficient $\alpha=0.15$ gives the most consistent band gaps for various iron oxides comparing with experimental ones \cite{He:prb14,Meng:JCTC,Tunega:JPCC}. With $\alpha=0.15$, the band gap of CaOFeS calculated using HSE06 is $1.63$ eV. Such a band gap remains suitable for absorption of solar light considering the photon energy of visible light (see the energy spectrum of solar light shown in Fig.~\ref{Fig6}(b)). Furthermore, the inherent polar structure may be advantaged for separation of photogenerated electron-hole pairs, which is also expected in ferroelectric materials \cite{Butler:EES,Yuan:Nm,Huang:Prb}.

Theoretically, the optical properties of a material can be described by the imaginary part of the dielectric constant ($\varepsilon_2(\omega)$), which can be obtained from the momentum matrix elements with the selection rules,
and the real part $\varepsilon_1(\omega)$ of the dielectric function can be calculated from imaginary part $\varepsilon_2(\omega)$ using Kramer-Kronig relationship ~\cite{Huang:Prb}. Then the absorption coefficient $\alpha(\omega)$ can also be derived. More details for these calculations can be found in the Appendix of Ref.~\cite{Huang:Prb}.

Due to the quasi-layer structure, the dielectric function is anisotropic: $\varepsilon_{xx}(\omega)=\varepsilon_{yy}(\omega)\neq\varepsilon_{zz}(\omega)$,
as shown in Fig.~\ref{Fig6}(a). The absorption coefficient $\alpha(\omega)$ is shown in Fig.~\ref{Fig6}(b). For $\varepsilon_2(\omega)$, there are several peaks in the visible light range. The first peak appears near $1.81$ eV (red light), corresponding to the first peak of the absorption coefficient $\alpha(\omega)$. It is natural that the crystal anisotropy leads to the anisotropy of the dielectric constant. Therefore, the absorption, which is a function of delectric constant, is also anisotropy.

Finally, the maximum photovoltaic energy conversion is estimated using a spectroscopic limited maximum efficiency method \cite{Yu:prl12}, as shown in Fig.~\ref{Fig6}(c). By taking the minimum direct gap  $1.81$ eV as $E_{g}^{\rm da}$ near Fermi level, the efficiency increases with the thickness of sample, whose maximum limit can reach $\sim24.2\%$. Compared with the estimated efficiency of some other photovoltaic materials, e.g. AgInTe$_2$($\sim27.6\%$), CuBiS$_2$($\sim16\%$), CH$_3$NH$_3$PbI$_3$ ($\sim30\%$) and CuBiS$_2$($\sim22\%$) ~\cite{Yu:prl12,Yu:AEM,Yin:AM14} (calculated using the identical method), this efficiency is still valuable. It should be noted that the polar effect has not been taken into account in the model, which enhance the electron-hole separation and thus improve the efficiency. In this sense, CaOFeS may be a potential photovoltaic material with prominent efficiency.

\section{Conclusion}
In summary, the physical properties of CaOFeS have been theoretically investigated. The G-type antiferromagnetic order has been confirmed to be the ground state by both the DFT as well as MC simulation. The G-type antiferromagnetism can break the trigonal symmetry of iron triangle lattice, and the exchange striction leads to magnetodielectric effect. Pyroelectrity is expected upon the antiferromagnetic transition, although the material is not ferroelectric. Furthermore, the large optical absorption coefficient has been predicted and the maximum photovoltaic energy conversion is estimated to be $\sim24.2\%$.

\acknowledgments{We thank Y. H. Li, M. F. Liu, and X. X. Chen for helpful discussions. This work was supported National Natural Science Foundation of China (Grant No. 11674055). Most calculations were done on Tianhe-2 at National Supercomputer Centre in Guangzhou (NSCC-GZ).}

\bibliographystyle{apsrev4-1}
\bibliography{ref3}
\end{document}